\documentclass[preprint,showpacs,tightenlines,preprintnumbers,amsmath,amssymb]{revtex4}

\usepackage[dvips]{graphicx}

\def\br{\bf r}

\begin{document}

\title{Evolution of deformations in medium-mass nuclei}

\author{H.  Sagawa,
   ~X. R. Zhou \footnote{present address: 
 Department of Physics, Xiamen University, Xiamen 361005, 
P. R. China} and X. Z. Zhang \footnote{On leave from China Institute of
Atomic Energy ,  Beijing, China}}

\affiliation{Center for Mathematical Sciences,  University of Aizu\\
Aizu-Wakamatsu, Fukushima 965-8560,  Japan\\
E-mail: sagawa@u-aizu.ac.jp}


\begin{abstract}
Evolution of quadrupole deformations in $sd$ and $pf$ shell nuclei with mass A=
18$\sim$56  is studied by using deformed 
 Skyrme Hartree-Fock (HF) model with pairing correlations.
    We point out  that the quadrupole deformations of
 the nuclei with the  isospin T=0 and   T=1 show 
   strong mass number dependence
as a clear manifestation of dynamical evolution of deformation in
nuclear many-body systems.  The competition between the deformation driving 
particle-vibration coupling and the closed shell structure is shown
in a systematic study of the ratios between the proton and neutron  deformations
in nuclei with T=$|$T$_z|$=1.  Calculated 
 quadrupole and hexadecapole deformations  are
   compared with  shell model results and 
 available experimental data. A relation between the skin thickness and the 
intrinsic Q$_2$ moments is also discussed.
\end{abstract}

\maketitle

\section{Introduction}
A study of the mass number 
 dependence of the deformations in nuclei is an interesting subject 
 in relation to a
 manifestation of spontaneous symmetry breaking  in quantum many-body 
systems.  
  The effect of spontaneous symmetry breaking 
is a general phenomenon known in many fields of physics.
In molecular physics,  the spontaneous symmetry breaking was discovered 
by Jahn and Teller in 1937~\cite{JT37}.
 The coupling to the quadrupole vibration is the main 
origin of the static deformation in both molecules and 
atomic nuclei~\cite{PG84}.
 The pairing correlation  in atomic nuclei  is known to  stabilize 
the spherical symmetry.  
An unique  feature of the evolution of deformation
  in atomic nuclei
will appear in 
the competition between the deformation deriving particle-vibration
coupling and the pairing correlations~\cite{Naz93}.
Top of that, the shell structure of  nuclei will give rise to 
a variety of different shapes, i.e., prolate, oblate and triaxial 
shapes depending on the position of the 
Fermi energy between two closed shells.

Electromagnetic observables  provide  useful information to
study structure of nuclei of both  ground states and  
excited states. Namely, the  observables  such as quadrupole (Q$_2$)
moments and electric quadrupole (E2) transitions are expected to pin down
 precise information  of deformations.    
The  evolution mechanism of deformation  will change the shapes of nuclei 
   from the beginning to  the end of the closed shell. 
Up to now,  the mass number  dependence of the deformations
 has been studied mainly 
in rare earth nuclei.  However, it is rather difficult in rare-earth and 
heavy nuclei to study the deformations  systematically 
between two closed shells due to the limitation of available nuclei for
experiments.  On the other hand, it is possible to access 
experimentally  from the beginning to the end of the closed shell 
in $sd$ and $pf$ shell nuclei.  Thus,  the   
$sd$ and $pf$ shell nuclei  might be  promising candidates
to discuss  the dynamical evolution of deformation 
 since several electric
quadrupole transitions are already observed \cite{Raman01} and 
new data of proton-rich unstable nuclei  are also available
 \cite{Yur04a,Yur04b,Yamada04}.

The main aim of this paper is to study the dynamical evolution of 
nuclear deformation in $sd$ and $pf$ shell nuclei
  with the isospin T=0 and 1 including unstable proton-rich nuclei.
To this end, 
 we perform deformed Skyrme Hartree-Fock (HF)+BCS  calculations
 with a density-dependent pairing interaction 
in nuclei of mass number   A=(16$\sim$56).   
Firstly, the mass number dependence of proton deformation is
 examined in comparison 
with empirical values extracted from the electric quadrupole 
transition strength B(E2) of  the T=0 nuclei. 
The quadrupole moments of the first excited 2$^+$ states are also 
studied  to extract the sign and the magnitude of the intrinsic 
quadrupole deformations.
 Then the
ratios  between calculated proton and neutron deformations of  T=1 nuclei 
are  compared with shell model results and    
 empirical values obtained from B(E2) values of mirror nuclei
  with  T=1 and  T$_z=\pm 1$, 
 assuming the isospin  symmetry in the two nuclei.
The evolution of higher multipole deformation 
is also an interesting subject 
 related with the shell filling effect of the deformed potential.
We will study the hexadecapole deformation by using the deformed HF wave 
function 
 and compared with available empirical data.  

 This paper is organized as follows.  The effect of the dynamical 
symmetry breaking on nuclear deformation
  is reviewed in Section IIA.  Deformed HF+BCS 
calculations  with Skyrme forces and  a density dependent 
pairing interactions are
presented in Section IIB.   The  calculated quadrupole moments 
  are compared with experimental
data in Section IIB and IIC.  
The isospin symmetry of the intrinsic 
Q$_2$  moments 
 is  discussed in comparison with
   shell model
 calculations and other mean field models in Sections IIIA and IIIB. 
A relation between the skin thickness and Q moments is  mentioned 
in Section IIIB.    Section IIIC is devoted to the study of  
hexadecapole deformations of the $sd-$ 
and $pf-$shell nuclei  calculated by  the HF wave functions and
compared with available experimental data.
  A summary is given in Section IV.

\section{Deformations in medium-heavy nuclei in mean field theory}
\subsection{Spontaneous symmetry breaking and pairing correlations}
  The phenomenon of spontaneous symmetry breaking  appears 
 in many field of physics.  
As an  example of the symmetry breaking, the deformation effect was 
 observed  commonly in quantum many-body systems,
 not only in molecules, but also atomic nuclei and atomic clusters.
 The spontaneous symmetry breaking was  recognized also  
 in the field theory,  superconductors and  condensed matter physics.
  Recently, it was discussed  in relation to 
the chiral band of tri-axial nuclei
~\cite{Koike04}.

Let us briefly discuss the  spontaneous symmetry breaking effect to
derive the nuclear deformation.  The simplest example is the case of
two degenerate single$-$particle states interacting with one collective 
mode denoted by the normal mode coordinate Q.
The collective hamiltonian for the  system can be given by the 
sum of the kinetic energy and the potential energy ,
\begin{equation}
 H(\mbox{Q})= -\frac{1}{2}\frac{\partial ^2}{\partial \mbox{Q}^2}
     +\frac{1}{2}\mbox{Q}^2 -k\mbox{Q}
  \left( \begin{array}{cc}
       1 &0 \\
    0 & -1 
    \end{array}
    \right)
\label{eq:coll}
\end{equation}
where the energy unit  of this hamiltonian (\ref{eq:coll}) is 
the eigenvalue of the normal mode.  
This collective hamiltonian is obtained based on the Born-Oppenheimer 
approximation, i.e., the decoupling between the collective 
and single particle motions.  This hamiltonian gives the potential minima
of $E=-k^2/2$  at both side of the deformation 
\begin{equation}
\mbox{Q}=\pm k  \,\,\,\,\, (k>0)
\label{eq:defm}
\end{equation}
 The positive value $+k $ gives a prolate deformation,
 while the negative value $-k $  gives an oblate deformation. 
The pairing correlations play an important role for the structure of 
not only  ground states but also  excited states in nuclei.
The interplay between the deformation driving force and the 
pairing correlations gives an unique feature of the symmetry breaking 
of atomic nuclei compared to other quantum many-body systems \cite{Naz93}.
We will discuss the effect of pairing correlations on 
the deformation later in this section and also in Section IIIA.

\begin{figure}[h]
\includegraphics[width=5in,height=5in]{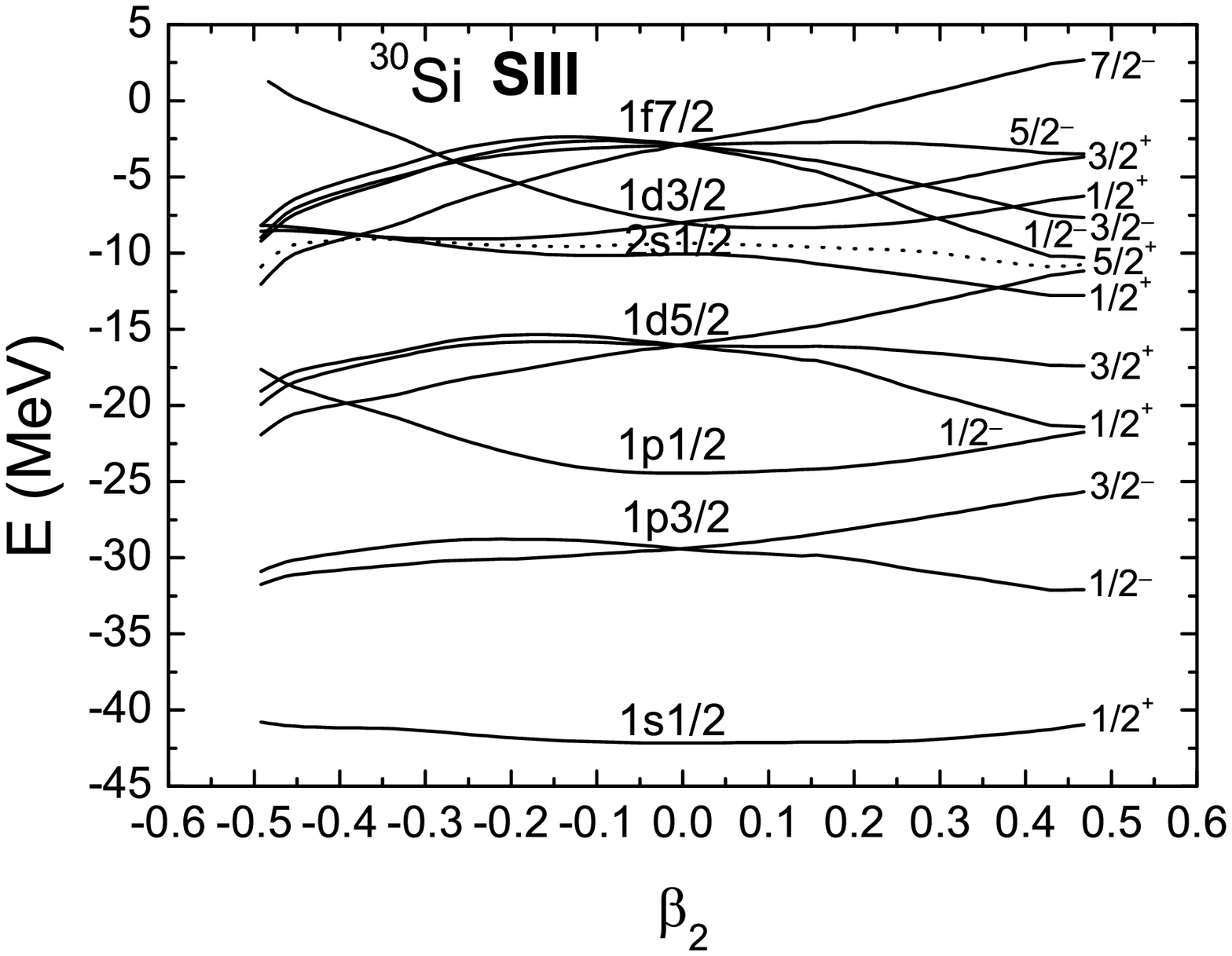}
\caption{\label{fig:nilsson30}
Neutron HF single particle energies as a function of the deformation
parameter $\beta_2$ in $^{30}$Si.  The deformed HF calculations are
performed by  using SIII interaction.}
 \end{figure}

 For a harmonic oscillator potential in the limit of large quantum numbers,
there is a symmetry between the prolate shape 
 occurring  at the beginning 
of the shell and the oblate shape at the end of the shell~\cite{BM75a}.
The HF neutron single-particle energies of 
$^{30}$Si  are given in Fig.~\ref{fig:nilsson30} as a function of the 
deformation parameter $\beta_2$.   
In $sd$ shell nuclei, 
 the strong preference for prolate deformation in the beginning of 
the shell can be understood from the fact that 
the energy of the orbit K$^{\pi}$=1/2$^+$ from $1d_{5/2}$ orbit (in the 
spherical limit) decreases more strongly as a function of deformation 
than do any  of the  orbits  in a  potential of oblate shape near the Fermi 
surface as is shown in Fig. \ref{fig:nilsson30}.
Namely the strength of the linear term in the hamiltonian (\ref{eq:coll})
is the largest for the prolate side at the beginning of the shell in 
 the HF  calculations.  
For the  harmonic oscillator potential, one would expect a preference of 
the oblate shape at the end of the shell due to the symmetry of the 
single$-$particle energies.  However, the spin$-$orbit force and the ${\vec
 l}^2$ term in the potential reduce the driving force for  oblate deformation 
 although the argument of the harmonic oscillator potential prefers it in the
latter half region  of  the closed shell.  Thus, 
in most of  nuclei, the calculated and 
observed deformations are very much dominated by the
 prolate shape \cite{Tajima01}.  However, in light nuclei like C and Ne 
 isotopes, it was pointed out that the deformation will change 
in a similar way to be expected from the deformed 
harmonic oscillator potential since the spin$-$orbit force and the ${\vec l}^2$
term are relatively small in the mean fields  of these isotopes \cite{Sag04}.

Higher multipole deformations are also associated with 
 the spontaneous symmetry breaking of nucleus.  The effect of the
shell filling in deformed potentials was discussed on the higher multipole 
deformations in refs. \cite{BM75a} and \cite{JZ84}.   We focus
on the hexadecapole deformation Q$_4$ in this paper and discuss the 
mass number dependence of it.

In the past, many theoretical model calculations were performed
 to study the masses and the deformations in $sd$ and $pf$ shell nuclei
 (See, for examples, review articles of various models in refs. 
 \cite{Quentin,Aberg}.
 Ragnarsson and Sheline carried out the Nilsson Strutinsky type 
 calculations with modified oscillator potential to discuss 
the deformation of nuclei in the whole nuclear periodic table  
\cite{Rag1,Rag2}.  
Moller and collaborators performed 
 very extensive calculations with 
a finite-range droplet model (FRDM)  
in the whole mass region from $^{16}$O to $^{339}$136~\cite{Moller}.
Various self-consistent models are also applied to study the medium mass 
nuclei; HF model  \cite{Flocard,Castel,JZ84}, HF+BCS model  \cite{Bonche},
HF+Bogoliubov model \cite{Girod,Goeke}.  Tri-axial rotor model was also used
to study the deformation in $^{30}$S \cite{Kurath72}.  

\subsection{Mass number dependence of deformations in T=0 nuclei}

We investigate  the  mass  number dependence of deformations
 in $sd$ and $pf$ shell nuclei with the isospin T=0 and T=1.  To this end, 
we perform deformed HF+BCS calculations with two Skyrme interactions
SGII and SIII to see the interaction dependence of the results. 
Both interactions are
commonly used in the mean field calculations and also random phase 
approximations for the excited states.  
The axial symmetry is assumed  for  the HF deformed
potential. The pairing interaction  is taken to be a density
dependent pairing interaction  in BCS approximation; \begin{equation}
V({\br_1},{\br_2}) =V_0' \left(1-\frac{\rho (r)}{\rho _0}\right)\,
\delta(\br_1-\br_2) \label{eq:pair_s} \end{equation} where $\rho (r)$ is the
HF density at ${\br} = ({\br}_1+{\br}_2)/2$ and $\rho _0$ is
chosen  to be 0.16 fm$^{-3}$. The pairing strength is taken to be
$V_0'=-410$ MeV$\cdot$fm$^3$ for both neutrons and
protons~\cite{DHF}. A smooth energy cut-off is employed in the BCS
calculations~\cite{BRRM00}.
We study in this section quadrupole 
  moments  Q$_2$ 
using the constrained deformed HF wave functions.   The constraint 
is imposed on the axial mass quadrupole moment operator \^{Q}$_{20}$.
 The intrinsic proton Q$_{2p}$ and neutron Q$_{2n}$ moments are
calculated by using HF wave function $|K>$, \begin{eqnarray}
\mbox{Q}_{2p}&=& <K|\int {\hat \rho}_p (x,y,z) (2z^2 -x^2 -y^2)d{\bf r}|K> 
\label{eq:q0p}  \\
\mbox{Q}_{2n} &=&<K|\int  {\hat \rho}_n (x,y,z) (2z^2 -x^2 -y^2)d{\bf
r}|K> 
\label{eq:q0n}
\end{eqnarray} 
where ${\hat \rho}_p$ and  ${\hat \rho}_n$ are the
proton and the neutron density operators.   
The intrinsic proton hexadecapole Q$_{4p}$ and neutron hexadecapole 
 Q$_{4n}$ moments are
also calculated by using HF wave function $|K>$, 
\begin{eqnarray}
\mbox{Q}_{4p}&=& <K|\int {\hat \rho}_p (x,y,z) (2z^4 -6z^2(x^2 +y^2)
+\frac{3}{4}(x^2 +y^2)^2)d{\bf r}|K> 
\label{eq:q4_p}  \\
\mbox{Q}_{4n} &=&<K|\int  {\hat \rho}_n (x,y,z)(2z^4 -6z^2(x^2 +y^2)
+\frac{3}{4}(x^2 +y^2)^2)d{\bf
r}|K> 
\label{eq:q4_n}
\end{eqnarray} 

The electric multipole transition strength B(E$\lambda$) 
   in the laboratory system can be expressed as 
\begin{equation}
B(E\lambda ;KI_i\rightarrow KI_f)=\frac{2\lambda +1}{16\pi}
 (e\mbox{Q}_{\lambda p})^2<I_iK\lambda 0|I_fK>^2
\label{eq:E2-moment} 
\end{equation} 
with the intrinsic Q$_{\lambda p}$ moment.  Furthermore, 
the quadrupole moment of the excited 2$^+$ state Q$_2(2^+$) 
can be  calculated by using the intrinsic quadrupole moment Q$_{2p}$;
\begin{equation}
\mbox{Q}_2(2^+)= -\frac{2}{7}\mbox{Q}_{2p}.
\label{eq:q2-moment}
\end{equation} 
The magnitude of  intrinsic Q$_{\lambda p}$ moments can be  
 extracted experimentally from observed   B(E$\lambda$) values  
by using Eq. (\ref{eq:E2-moment}).   
 The observed  Q$_2(2^+$) moment  will be useful
to extract not only  the magnitude but also the sign of the intrinsic
 Q$_{2p}$ moment from Eq. (\ref{eq:q2-moment}).
 
\begin{figure}[p]
\includegraphics[width=16cm,clip]{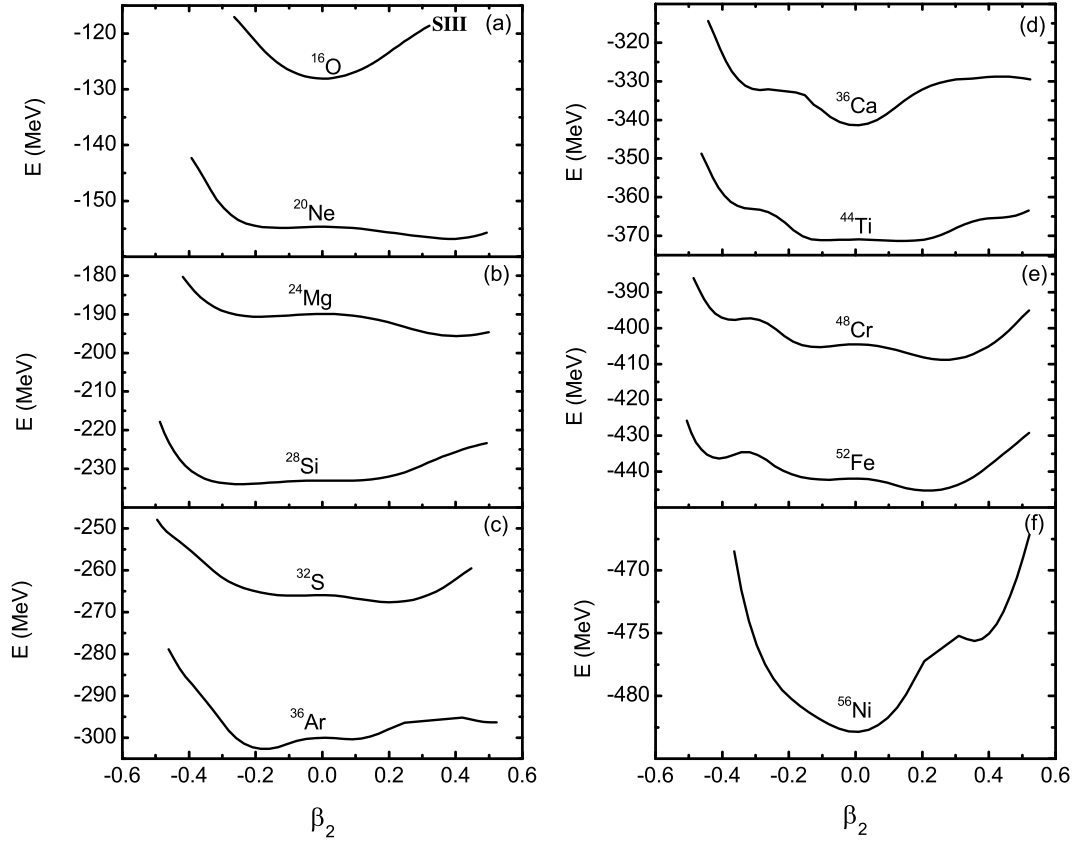}
\caption{\label{fig:energy_T0}
Deformed HF $+$  BCS calculations of T=0 nuclei  with SIII interaction;
 (a) for $^{16}$O and $^{20}$Ne,  (b) for  $^{24}$ Mg and $^{28}$Si , 
 (c) for  $^{32}$S and $^{36}$Ar, (d) for  $^{36}$Ca and $^{44}$Ti, 
 (e) for  $^{48}$Cr and $^{52}$Fe and  (f) for  $^{56}$Ni.  
 The density dependent pairing
interaction (\ref{eq:pair_s}) is adopted in the calculations.}
\end{figure}

  In Fig. \ref{fig:energy_T0},  the calculated HF 
energy surfaces of T$=0$ nuclei with SIII interaction 
are shown for nuclei with the mass A= 16$\sim$56.  
The proton, neutron and mass deformation 
parameters  $\beta_{2p}$, $\beta_{2n}$ and $\beta_{2}$  at the minimum of
the energy surface are defined by
\begin{eqnarray}
\beta_{2p}=\sqrt{\frac{\pi}{5}}\frac{\mbox{Q}_{2p}}{Z<r^2>_p} 
       \label{eq:q_betap} \\
\beta_{2n}=\sqrt{\frac{\pi}{5}}\frac{\mbox{Q}_{2n}}{N<r^2>_n} 
 \label{eq:q_betan}\\
\beta_{2}=\sqrt{\frac{\pi}{5}}\frac{\mbox{Q}_{2}}{A<r^2>_m}
\label{eq:q_beta}
\end{eqnarray}
where $<r^2>_p, <r^2>_n$ and $<r^2>_m$  are  the proton, neutron and mass 
mean square radius, respectively.  The calculated deformation 
parameters are 
tabulated in Table~\ref{tab:Q-T0} for SIII and SGII 
interactions together 
 with the intrinsic proton and
neutron quadrupole moments Q$_{2p}$ and Q$_{2n}$. 
  The shape isomers are also 
found in several nuclei at low energies, which are also tabulated in
Table~\ref{tab:Q-T0}.  
 The sign of deformation is always the same for
protons and neutrons in a given configuration of each nucleus, 
  while the magnitude is largely different in
several configurations.

An interplay between the single particle
energy gap  around the Fermi surface and the quadrupole vibration energy 
plays an essential role to drive the deformation.
As is expected from the large energy gaps at  the closed
shells  with N=Z=8 and  N=Z=20, the doubly closed shell nuclei 
 $^{16}$O and $^{40}$Ca show spherical shapes.  The prolate deformations grow in  $^{20}$Ne
 and  $^{24}$Mg. In the middle of the two closed shells,
  the prolate and the oblate shapes are competing  in $^{28}$Si and 
$^{32}$S so that the 
energy surfaces become rather flat.  Just before the end of the closed shell 
 N=Z=20, the oblate shape appears as the ground state configuration 
of $^{36}$Ar.  Above the mass A=40, the prolate deformations are dominant
 in  $^{44}$Ti, 
$^{48}$Cr and  $^{52}$Fe.  Then the shape becomes again 
  spherical in a closed shell 
nucleus $^{56}$Ni.     The  local oblate minima are
found at low excited energies in the three nuclei  A=44,48 and 52 . 
Since the slopes of K=1/2$^-$ and K=3/2$^{-}$ states from $f_{7/2}$ orbit 
  go   downward  steeper
  in the the prolate side,   as can be seen in Fig.~\ref{fig:nilsson30}, 
  the  prolate deformations are favored in the ground states of these nuclei.
 On the other hand,  in the nuclei A=28, 32 and 38, 
 the whole $sd$ shell orbits 
are involved and give the competition between oblate and prolate deformations.
In $^{44}$Ti, it is found that 
the prolate and the oblate minima are almost degenerate 
in energy.  However the two energy minima are very shallow and have small 
deformations so that the dynamical fluctuations will dominate low 
excitation energy spectra and 
it is unlikely to have any rotational bands. 

\begin{figure}[p]
\includegraphics[width=4in,height=5in]{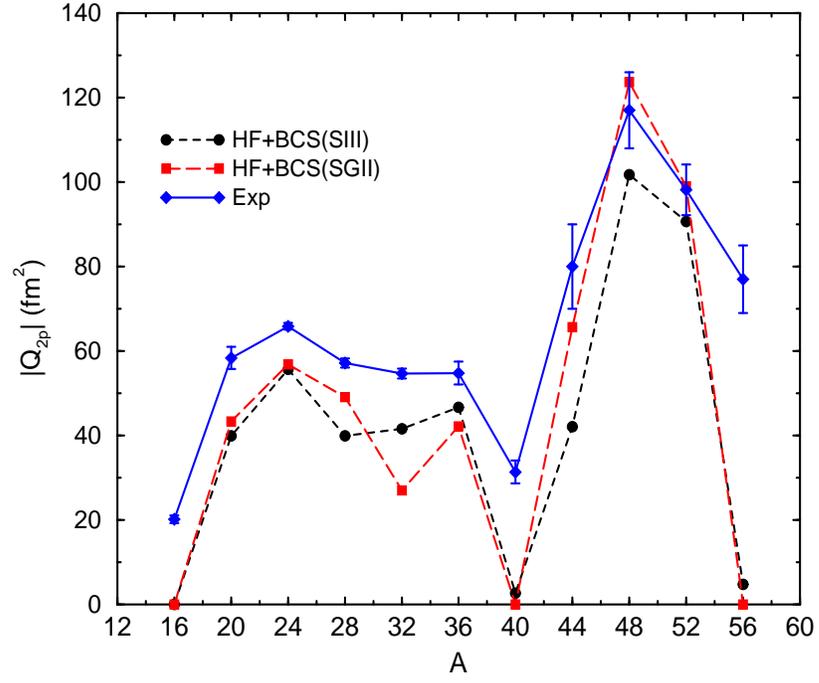}
\hspace{-2cm}
\caption{\label{fig:q0_tz0}
Absolute values of 
 proton quadrupole moments in $sd$ and $pf$ shell nuclei with N=Z. 
Deformed HF $+$  BCS calculations are performed  with SIII and SGII
 interactions.   The density dependent pairing
interaction (\ref{eq:pair_s}) is adopted in the calculations.
The experimental data are taken from refs. \cite{Raman01,Yamada04}.}
\end{figure}

The calculated Q$_{2p}$ moments are compared with experimental data in
Fig.~\ref{fig:q0_tz0}. 
The experimental Q$_{2p}$ moments are extracted from the transition strength
$B(E2)$  between the ground states and the first excited 2$^+$ states
 by using Eq.~(\ref{eq:E2-moment}).
 A clear manifestation of dynamical 
evolution of deformation can be seen 
   in both the theoretical and experimental Q$_{2p}$ 
moments.  Namely, the  Q$_{2p}$ moment is  very small in the 
closed shell nucleus with A=16. Then it increases
   when few particles are added to the 
 closed core  and shows  a kind of plateau at the middle 
 of the shells.  A similar pattern is seen in the nuclei between A=40 and 56.
 Quantitatively, the calculated Q$_{2p}$ 
 moments are about 10$\sim $20\% smaller
than the empirical one in nuclei with A=20$-$36.  However, the agreement is 
 better in $pf$ shell nuclei.  Especially SGII interaction gives almost 
 identical results to the empirical ones of nuclei with 
A=44$-$52 within the statistical uncertainties.

It is shown that the calculated deformations of the ground states 
 $^{28}$Si and $^{36}$Ar are oblate 
  shapes. In both nuclei, the slopes of deformed HF 
  single-particle energies in Fig.~\ref{fig:nilsson30} favor 
the oblate deformations.
 As a result, in $^{28}$Si, the oblate shape is induced by the occupation of 
K$^{\pi}$=1/2$^{+}$ state coming from 1d$_{5/2}$ state in the spherical limit,
   while the K$^{\pi}$=5/2$^{+}$
 state is going up rapidly in the prolate side.  
  The oblate shape in  $^{36}$Ar 
is  traced back to the occupation of the K$^{\pi}$=3/2$^{+}$ state
 coming from 1d$_{3/2}$ state in the spherical limit as well.

Deformed nuclei in the $sd$ shell have studied with various theoretical
 models.  As far as the quadrupole deformations of 
N=Z nuclei with A=(20$\sim $36) are concerned, the present results are
 very similar to the previous calculations by the HF models 
with various interactions \cite{Quentin,JZ84} 
and the modified Nilsson model ~\cite{Rag2}.  
 Namely,  $^{20}$Ne,  $^{24}$Mg and $^{32}$Mg were found to be prolate 
 in the ground states, while  $^{28}$Si and $^{36}$Ar are oblate. 
 Thus, the evolution mechanism of the deformation mentioned here is 
 common in the various microscopic models.

\subsection{Mass number dependence of deformations in T=1 nuclei}

\begin{figure}[p]
\includegraphics[width=16cm,clip]{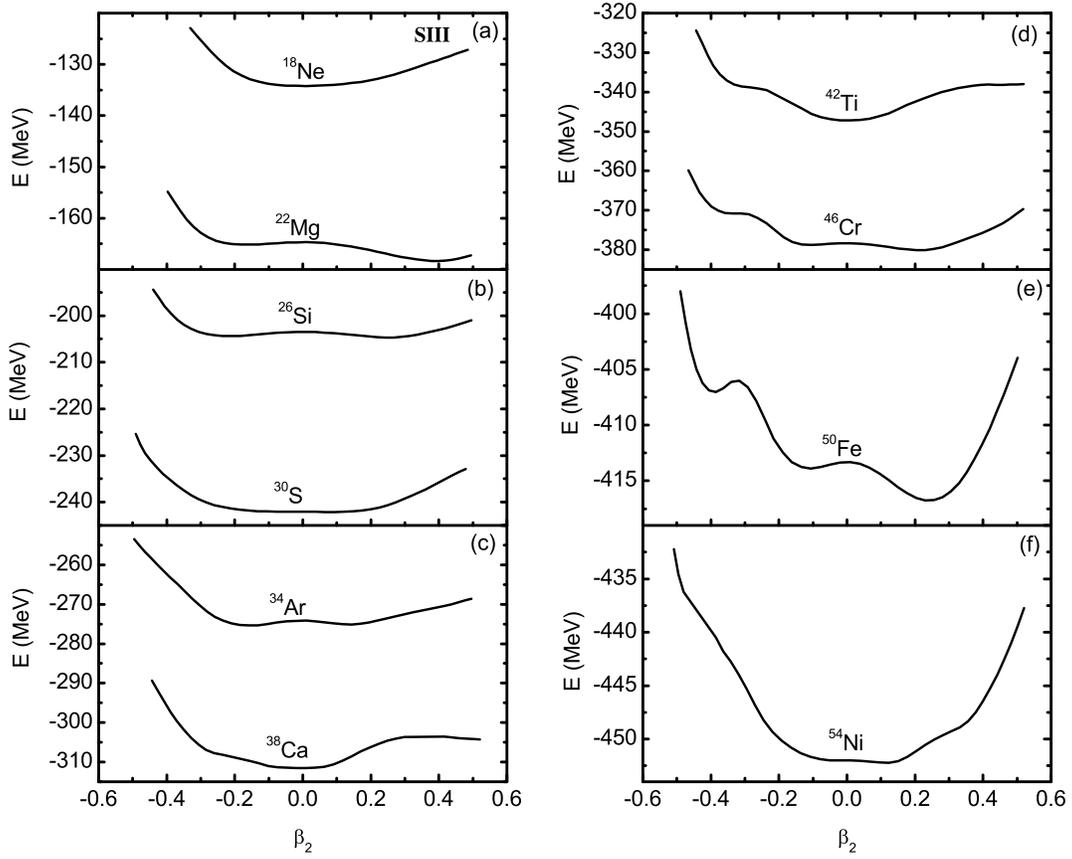}
\caption{\label{fig:energy_T1}
Deformed HF $+$  BCS calculations of T=1 with SIII interaction; (a)
 for  $^{18}$Ne and $^{22}$Mg,   (b) for  $^{26}$Si and $^{30}$S,  
 (c) for  $^{34}$Ar and $^{38}$Ca,   (d) for  $^{42}$Ti and $^{46}$Cr,   
(e) for  $^{50}$Fe and  (f) for  $^{54}$Ni.  
The density dependent pairing
interaction (\ref{eq:pair_s}) is adopted in the calculations. }
\end{figure}

Fig.~\ref{fig:energy_T1}   shows the binding energy surfaces  of nuclei with
 T$=1$ with T$_z  =-$1.   The proton, neutron and mass deformation 
parameters  $\beta_{2p}$, $\beta_{2n}$ and $\beta_{2}$  at the minimum of
the energy surface are tabulated
  in Table~\ref{tab:Q-T1} for SIII and SGII interactions 
together with the proton and
neutron quadrupole moments Q$_{2p}$ and Q$_{2n}$.
  To study the mirror symmetry of deformations,
  the results of  nuclei with T$=1$ with T$_z$=1 are also listed in  Table
~\ref{tab:Q-TZm1} for SIII and SGII interactions.  
Since it is known that
 the correlations beyond the mean field approximations are 
important in nuclei near  $^{16}$O and $^{40}$Ca \cite{Lang01,Bender},  
the results of nuclei with A=18 and 40$\pm $2 are shown to  illustrate  
 the systematic trend of the deformation from the beginning to the
 end of the shells as a manifestation of the nuclear Jahn-Teller effect.
 The large prolate  deformations  
 are found in A=22 nuclei $^{22}$Mg and $^{22}$Ne with $\beta_2 \sim$0.4.
The calculated proton quadrupole moment
  Q$_{2p} \sim $50 fm$^2$ (43  fm$^2$) for $^{22}$Mg ($^{22}$Ne)  shows good
agreement  with the empirical value Q$_{2p}$= 60$\pm$11 fm$^2$ 
(48.1 $\pm$1.0 fm$^2$ )~\cite{Raman01}.   The calculated  shapes are 
  prolate in nuclei
 with smaller mass than the middle of 
  the two shell closures, i.e., in  $^{22}$Mg and $^{26}$Si.
 The prolate and oblate minima  appear at almost the same energy in 
   $^{26}$Si and $^{26}$Mg, while the energy surface is very flat in the
case of $^{30}$S as can be seen in Fig.~\ref{fig:energy_T1}(b).
In $^{34}$Ar and  $^{34}$S having heavier  mass than the middle
 of the two closed shells,
  the oblate minima are  found 
 slightly lower than the prolate ones in energy.  
  In the $pf$ shell nuclei, the prolate shapes 
 dominate in the nuclei with A=46, 50 and 54 as seen 
 in Tables~\ref{tab:Q-T1}, 
  and \ref{tab:Q-TZm1}.

Fig.~\ref{fig:q_tz1} shows the ratio Q$_{2p}$/Q$_{2n}$ divided by 
 the ratio of proton 
 to neutron number  Z/N in nuclei with T=1 and  T$_z=-$1. 
Because of strong proton$-$neutron 
interaction in the mean field,  the value (Q$_{2p}$/Q$_{2n})$/(Z/N)
 becomes close to unity when the
deformation  is well developed.
The empirical neutron deformations in the nuclei with  T=1 T$_z=-$1 
  are  extracted from the proton deformations  in the corresponding
  T=1 T$_z$=1  nuclei with the same mass  
 assuming a mirror symmetry of the deformations between the two nuclei. 
 As far as the deformed HF calculations are concerned,
  the mirror symmetry of deformation   
 is   well conserved in the nuclei of  A=(18$-$54)
 as seen  in comparisons with Tables ~\ref{tab:Q-T1} and 
 \ref{tab:Q-TZm1}.   
 The ratio (Q$_{2p}$/Q$_{2n})/$(Z/N) deviates largely  from unity 
  near the doubly closed shells  
 A=16 and 40.  Namely the ratio is more than 2 in A=18 system both in 
the HF calculations and the experiments.   Since  
  the driving force
 of deformation is only two protons in $^{18}$Ne,  
  a small proton deformation is induced.  
While there is no neutron particles 
which derive the neutron deformation in $^{18}$Ne, 
    the strong proton-neutron interaction 
 raises  smaller amount of  deformation  also
  for neutrons.  The same trend can be seen 
   in the nucleus with $^{42}$Ti having  two protons top of the 
doubly closed shell nucleus $^{40}$Ca.  
  The ratio  (Q$_{2p}$/Q$_{2n})$/(Z/N) becomes 
  opposite in the case of A=38 where 
  two neutron holes create a driving force for the deformation  so that 
  the neutron deformation is larger than the proton one although the 
  absolute magnitude is   small.  

 To confirm   theoretical conjectures  of the oblate deformations, 
it is necessary to obtain  
 empirical information of the sign of the quadrupole deformation. 
   To this end, 
 experimental data  of the quadrupole moment of excited 2$^+$ 
state Q$_2$(2$^+$) will  be 
 useful to determine not only the magnitude but also the sign of the 
  intrinsic quadrupole deformation with  Eq. (\ref{eq:q2-moment}).  
The present HF results of Q$_2$(2$^+$) are compared with 
experimental data and shell model calculations in Table \ref{tab:q2-moment}.
In general, the HF results show good agreement  with the observed data 
in both the magnitude and the sign
except the two nuclei $^{18}$O and $^{42}$Ca in which the correlations beyond
the mean field approximation are important \cite{Bender}.
It is seen that 
the observed data show clear evidence of the oblate deformations in
  $^{28}$Si and $^{36}$Ar. The shell model  calculations are also
 given  in Table 
 \ref{tab:q2-moment}.  In $sd$ shell nuclei, the shell model results are
very close to the HF ones, while we notice some differences between the 
two models in the $pf$ shell nuclei.
It is remarkable that our HF results give equally good or slightly better 
results in the   $pf$ shell nuclei compared with  modern version of 
shell model calculations.  

The two interactions give almost equivalent  results in Table 
\ref{tab:q2-moment} except $^{30}$Si  
in which even the sign is different. 
The  Q$_2$(2$^+$) moments in $^{30}$Si and $^{32}$S were discussed 
in many mean field models \cite{Castel,Goeke,Quentin} 
 and a triaxial rotor model 
\cite{Kurath72}.
  The HF and HF+Bogoliubov calculations with different interactions 
 gave large positive (oblate)  Q$_2$(2$^+$) in $^{30}$Si as was the same in  
the triaxial rotor model. One exception was 
the  SII interaction which  gave negative (prolate) value \cite{JZ84}. 
In Table \ref{tab:q2-moment}, the SIII gives a negative  Q$_2$(2$^+$) 
which is close to the shell model value, while the SGII shows  a small 
positive value.  At the moment, the accuracy  of the
 experimental data is not good enough to distinguish which interaction
 or which model is better or not to predict the Q$_2$(2$^+$) moment
 in $^{30}$Si.  

\section{Discussions}

\subsection{Mean field models and shell models}

An extensive study of deformed HF+BCS calculations was performed in ref.
 \cite{Tajima} by using SIII interaction with a seniority pairing interaction.
The results in ref.
 \cite{Tajima} are quite similar to the present ones as far as the deformations 
of the ground states of $sd$ shell nuclei A=16$\sim$40 are concerned.
 On the other hand, in T=0 nuclei of $pf$ shell,  the results of the two 
calculations are different qualitatively and also quantitatively with   
 the same Skyrme  interaction SIII.  In $^{44}$Ti, the present 
results with the surface $\delta $ pairing interaction 
 show a finite prolate deformation 
with $\beta_2$=0.12 for SIII interaction and $\beta_2$=0.19 for SGII interaction, 
while the result in ref. \cite{Tajima} shows no sign of deformation.  For 
Q$_{2p}$ and Q$_{2n}$ values in $^{48}$Cr and $^{52}$Fe, 
 the present results are (20$\sim$
30)\% larger than those of ref. \cite{Tajima} and close to the empirical values 
as shown in Fig.~\ref{fig:q0_tz0}.
For $pf$ shell nuclei with  T=1 , 
 the present results show finite prolate deformations
$\beta_2 \sim$0.2 for $^{46}$Cr and  $\beta_2 \sim$0.12 for $^{54}$Ni, while 
the results of ref.  \cite{Tajima} are spherical in both cases.  The Q$_2$
   moments 
of the present results in $^{50}$Fe are about 25\% larger than that of 
ref.  \cite{Tajima} and show
  better  agreement with experimental data.  

\begin{figure}[p]
\includegraphics[width=16cm,clip]{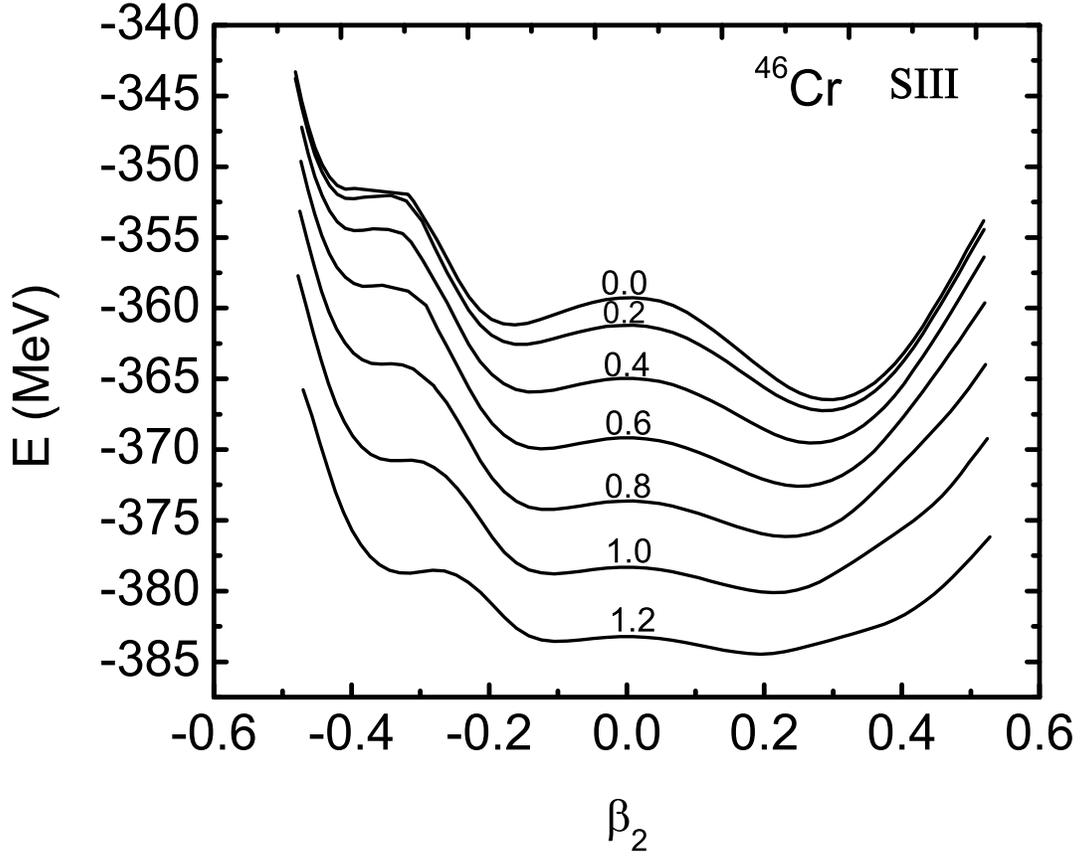}
\caption{\label{fig:pair_def}
Deformed HF $+$  BCS calculations of $^{46}$Cr with different  
pairing strength multiplying a factor 0.0$\sim$1.2 to the 
surface type pairing interaction (\ref{eq:pair_s}).
The SIII interaction is used.}
\end{figure}

 The magnitude of the quadrupole deformation depends on the strength of the
pairing interaction.  
In Fig. \ref{fig:pair_def},  the energy surfaces of $^{46}$Cr are shown 
changing the pairing strength 
multiplying a factor 0.0$\sim $1.2 by the surface type pairing interaction
(\ref{eq:pair_s}).  Without pairing correlation, 
 the energy minimum appears 
at $\beta_2\sim$0.3.  The energy minimum becomes shallower having smaller 
 $\beta_2$  when the pairing strength is increased.  
The adopted pairing strength (in the case of the factor 1.0) 
 in this study gives the minimum at $\beta_2$=0.2. As is expected, 
the binding energy becomes larger with stronger pairing interaction.
The pairing gain energy is about -16MeV in $^{46}$Cr and 
  the calculated total binding energies 
with the adopted pairing (\ref{eq:pair_s}) are E=-380MeV
 in $^{46}$Cr
 which is 
  close to the empirical one E(exp)=-382MeV.
As far as the pairing gain energy is concerned, 
the pairing correlation  is about 20\% stronger in 
 ref. \cite{Tajima} which gives in general  smaller deformations 
 in $pf$ shell nuclei and  predicts  spherical shapes in some nuclei like 
  $^{46}$Cr and $^{52}$Fe.

 The Skyrme HF calculations of T=0 nuclei with A=32$\sim$48 were performed 
 in ref.\cite{Inakura} on the symmetry unrestricted basis without 
paring correlations.  They found  a  similar isotope 
dependence of the quadrupole deformations  to that of the present 
calculations with the pairing correlations.  They pointed out also
that the octupole deformation is very small in the ground state 
configurations of the T=0 nuclei.

\begin{figure}[p]
\includegraphics[width=4in,height=5in]{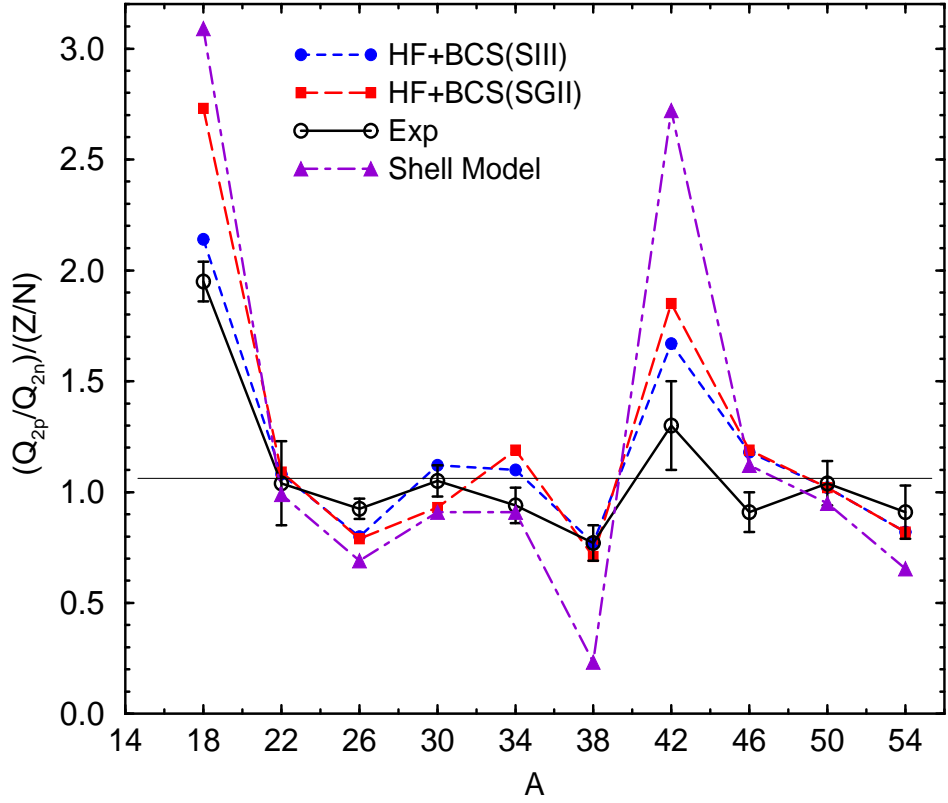}
\caption{\label{fig:q_tz1}
Ratio of proton to neutron quadrupole moments (Q$_{2p}$/Q$_{2n}$) 
  divided by that  of 
proton to neutron numbers (Z/N) 
 in $sd$ and $pf$ shell nuclei with T=1 and  T$_z=-1$. 
Deformed HF $+$  BCS calculations are performed  with SIII and SGII
 interactions.   The density dependent pairing
interaction (\ref{eq:pair_s}) is adopted in the HF calculations.
The shell model values  
are  calculated by using Eq. (\ref{eq:q_shell}). 
 The shell model transition matrices of $sd$ shell 
  are taken from ref.~\cite{BW82}, 
while those of $pf$ shell are taken from ref.~\cite{Honma04}.
The empirical values are obtained by assuming a mirror symmetry
between proton and neutron quadrupole moments in T=1 and  T$_z=\pm1$ nuclei
with the same mass number A.
Experimental data are taken from refs. \cite{Raman01,Yamada04}.
See the text for details.}  
\end{figure}

The proton and neutron transition matrix elements M$_p$, M$_n$ were calculated 
by the shell models in $sd$ shell nuclei by Brown and Wildenthal 
\cite{BW82} and in $pf$ shell nuclei by Honma et al.,\cite{Honma04}.
 The ratio of the Q moments
 (Q$_{2p}$/Q$_{2n})$ of  the shell models is obtained  by using 
 an equation with the values M$_p$, M$_n$  of T=1, T$_z=-1$ nuclei,
\begin{equation}
 (\mbox{Q}_p/\mbox{Q}_n)=
   (e^{eff}_p\mbox{M}_p+e^{eff}_n\mbox{M}_n)/(e^{eff}_p\mbox{M}_n+
  e^{eff}_n\mbox{M}_p),
\label{eq:q_shell}
\end{equation} 
which corresponds to the ratio of transition matrices between
 the mirror nuclei with 
T=1, T$_z=-1$ and T=1, T$_z=1$. 
The effective charges are taken to be $e^{eff}_p$=1.35 and $e^{eff}_n$=0.35
for $sd$ shell nuclei,  and  $e^{eff}_p$=1.5 and $e^{eff}_n$=0.5 
for $pf$ shell nuclei. 
In general, the mass number dependence of the 
shell model values (Q$_{2p}$/Q$_{2n})$/(Z/N) in 
 Fig.~\ref{fig:q_tz1}  is  similar to that 
 of the HF results. Namely, in $sd$ shell, the absolute values 
 (Q$_{2p}$/Q$_{2n})$/(Z/N) of
 the shell model calculations
are close    to those of HF calculations  
and show good agreement with empirical data except the nuclei with A=18 and 38.
 The shell model results deviate
 from unity  much larger than those  of HF  and also  empirical ones 
in the nuclei near the closed shells N=Z=8, N=Z=20 and N=Z=28 as  seen  
  in Fig.~\ref{fig:q_tz1}.  
The shell model 
  transition matrix elements depend largely  on the adopted model space
for  nuclei near the closed shell.  In the case of A=18, 
 a complete basis of $(1p_{1/2}, 1d_{1/2}, 2s_{1/2})$  model space
(so called ZBM basis) \cite{ZBM} 
gives a better result for 
the value  (Q$_{2p}$/Q$_{2n})$/(Z/N) to be 
  1.67, instead of more than 3 in ref.~\cite{BW82}, 
 in comparison with the empirical data.

In $pf$ shell nuclei, the shell model results are close to
 those of HF calculations  and consistent with the empirical values in
 the nuclei with A=46 and 50.  On the other hand, the shell model values are 
  much larger  in A=42 and smaller in A=54 in comparison with 
the HF results.  
These  results suggest that  a strong  proton$-$neutron interaction in the HF 
calculations induces  smaller asymmetries  in the deformations between 
T=1, T$_z=\pm$1 nuclei  near the closed shells than expected from the
shell model results.  In the middle of the closed shells, both the
 HF and shell model calculations give reasonable collective 
transition strength compared with the experimental values.

Finite-range droplet model (FRDM) has been known to provide 
useful information on the masses and the deformations of nuclei
in a very wide region of the  mass table ~\cite{Moller}.  
Compared with the HF model, the FRDM gives quite different results for the
quadrupole deformations in many $sd$ and $pf$ shell nuclei.   
For examples, the FRDM gives large deformations 
$\beta_2$=(0.32$\sim$0.37) for $^{20,22}$Ne and
$^{22,24}$Mg which are consistent with the HF results in Tables 
\ref{tab:Q-T0},\ref{tab:Q-T1},\ref{tab:Q-TZm1}.
A nucleus $^{26}$Mg is predicted
 as an oblate shape in FRDM, while the oblate and
 the prolate minima are almost degenerate in the HF results in Table ~\ref{tab:Q-TZm1}.
The S, Ar,  Cr and Fe isotopes show substantial deformations in 
the HF calculations, while these isotopes are predicted to be 
spherical in the FRDM. It seems that the FRDM might not give 
 proper deformations in these medium mass nuclei in comparison  with 
 empirical information discussed in this paper.

\subsection{New Experimental Data and Comparisons with Theories}

Recently, the B(E2) values of several proton-rich 
 unstable $pf$ shell nuclei with T=1 and T$_z$=$-$1
 are observed by Coulomb excitations \cite{Yur04a,Yur04b,Yamada04}.
The mirror symmetry in  T=1 nuclei with A=46,50 and 54 is experimentally 
confirmed by these new data.  Namely, in A=46, the 
  experimental values are  B(E2)=(950$\pm $50) $e^2fm^4$ in $^{46}$Ti   
and  (929 $\pm $199) $e^2fm^4$ in $^{46}$Cr, while 
 the calculated values are  B(E2)=582.1$e^2fm^4$ and 913.7
 $e^2fm^4$ for $^{46}$Ti and $^{46}$Cr, respectively, with SGII interaction.
The different calculated Q moments of the two nuclei 
are due to the proton skin effect in $^{46}$Cr which enhances the value
as is expected from Eq.~\ref{eq:q_betap}.
Experimental data of  T=1 and T$_z$=$-$1 nuclei with A=50 and 54 were reported 
to be B(E2)=(1359 $\pm $261) $e^2fm^4$ for $^{50}$Fe and
  B(E2)=(590$\pm $168) $e^2fm^4$
for $^{54}$Ni in ref. \cite{Yamada04}.  Corresponding experimental values for 
 T=1 and T$_z$=1 nuclei are B(E2)=(1080 $\pm $60) $e^2fm^4$ for $^{50}$Cr 
and B(E2)=(620 $\pm $50) $e^2fm^4$ for $^{54}$Fe \cite{Raman01}, respectively.
Calculated HF values of A=50 nuclei with SGII interaction,  
   B(E2)=1160$e^2fm^4$ for $^{50}$Fe and 1097$e^2fm^4$ for $^{50}$Cr,   
    show a clear mirror symmetry 
and give good accounts of  the 
experimental  data.  
The mirror symmetry is also seen  in the  HF calculations 
 of  A=54 nuclei with  B(E2)=373$e^2fm^4$ for $^{54}$Ni and 
390$e^2fm^4$ for $^{54}$Fe,  although the absolute values are smaller 
than the observed ones.  
In refs.  \cite{Yur04a,Yur04b}, the B(E2) values for the ground states 
  to the first 2$^+$ states were observed by the intermediate Coulomb 
  excitations which give almost identical results given in 
 refs. \cite{Raman01} and \cite{Yamada04}; 
B(E2)=(626$\pm $169) $e^2fm^4$ for $^{54}$Ni and 
 (640 $\pm $13) $e^2fm^4$ for $^{54}$Fe.

Shell model calculations of  B(E2) strength from the ground states to the
first excited states of A=50 and 54 nuclei were performed in a restricted 
 $pf$ shell model space with GXPF1 effective interaction \cite{Honma04}.
Taking the effective charges $e_p$=1.5 and  $e_n$=0.5, the shell model 
calculations give  a good mirror symmetry in the case of  A=50 nuclei
with B(E2)= 910  $e^2fm^4$ for $^{50}$Cr and 
 B(E2)=954$e^2fm^4$ for $^{50}$Fe although the absolute values are 
somewhat smaller than  the experimental values and those of deformed 
HF calculations with SGII interaction.  The shell model gives
a reasonable B(E2) value for $^{54}$Fe to be  B(E2)=651$e^2fm^4$ 
, while the calculated value is smaller than the experimental one 
in the case of  $^{54}$Ni as B(E2)=324$e^2fm^4$.  

\begin{figure}[p]
\includegraphics[width=4in,height=5in]{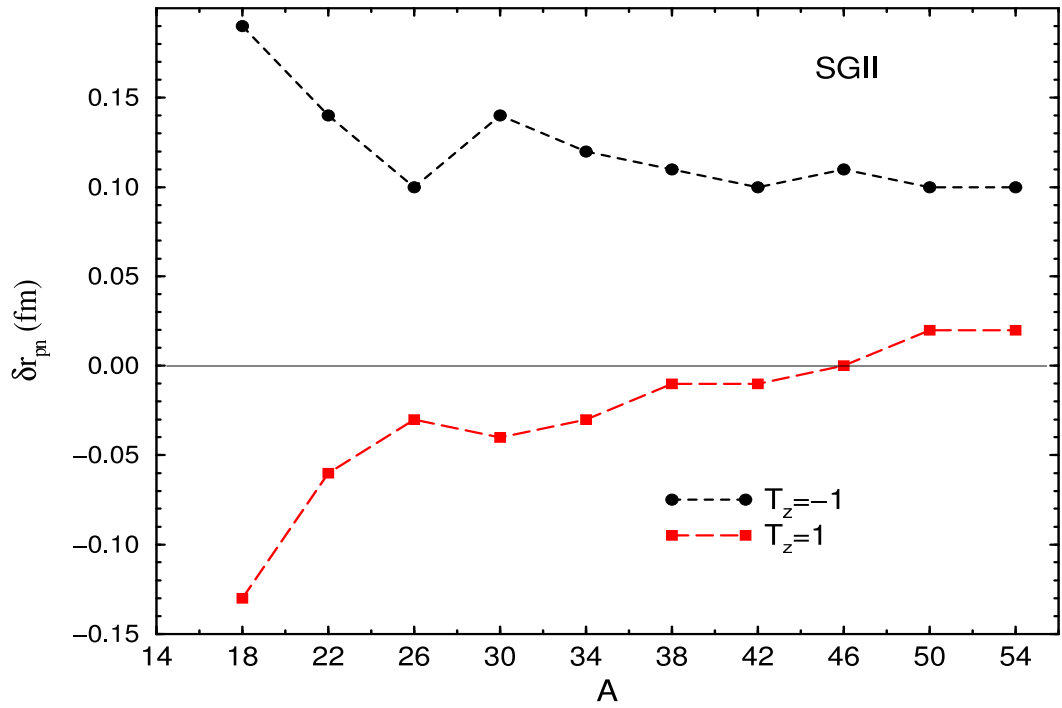}
\caption{\label{fig:radius_pn}
The difference between proton  and neutron rms radii
 defined by Eq. (\ref{eq:radius_pn}).  The SGII interaction is used
in the deformed HF calculations with a surface pairing interaction
 (\ref{eq:pair_s}) .
See the text for details.}  
\end{figure}

If the mean square radii are close  
$<r^2>_p \simeq <r^2>_n$ in a nucleus, the ratio Q$_{2p}$/Q$_{2n}$ divided by Z/N 
should be the same as the ratio $\delta_p /\delta_n$ (or equivalently 
$\beta_p/\beta_n$).
One can notice that T=0 nuclei in Table~\ref{tab:Q-T0}
show the proportionality between  Q moments and the deformation parameters 
$\beta_2$ as is expected from Eqs. \ref{eq:q_betap} and \ref{eq:q_betan}.  
  On the other hand, 
in Tables ~\ref{tab:Q-T1} and 
 \ref{tab:Q-TZm1}, the Q moments are not always proportional to
  the $\beta _p$ and $\beta _n$values.
    This is due to a difference of 
the proton skin effect in the T$_z=-1$ nuclei to the  
neutron skin effect  in the T$_z=1$ nuclei.  In Fig.~\ref{fig:radius_pn},
the calculated proton skin thickness of the ground state,
\begin{equation}
\delta r_{pn}=<r_p> - <r_n>,
\label{eq:radius_pn}
\end{equation}
is shown for the nuclei in $sd$ and $pf$ shell with T=1 and T$_z=\pm$1.
The results of SGII are shown in Fig.~\ref{fig:radius_pn}.  We 
calculated also the skin thickness by using the  SIII interaction and found
   almost identical results to those shown in Fig.~\ref{fig:radius_pn}.
In A=18, the proton-rich nucleus $^{18}$Ne has a large proton skin and 
the neutron-rich nucleus $^{18}$O  has an equally large neutron skin.
The  large skin thickness is created by two particle occupation of  the 
$1d_{5/2}$ orbit which has much larger radius than $1p$ orbits.
The proton skin thickness of nuclei with  T$_z=-$1 decreases gradually
for heavier $pf$ shell nuclei, but still finite to be 
$\delta r_{pn}\simeq 0.1$ even in $^{54}$Ni.
The irregularity at A=30 is due to the occupation of $2s_{1/2}$ orbit.
The neutron skin thickness of nuclei with  T$_z=$1   decreases more rapidly 
for  heavier $sd$ shell nuclei and almost disappears in the $pf$ shell nuclei.
This difference between the nuclei with T$_z=\pm$1 
is due to the effect of the Coulomb interaction on the 
mean field proton potential.  
Since the Coulomb potential makes shallower proton HF potential in $pf$ shell 
nuclei, the extra two protons give always the enhancement of the proton rms 
radii  in  T$_z=-$1 nuclei. On the other hand, 
 the neutron potential is deeper than the 
proton one in  T$_z=$1 nuclei so that the extra two neutrons do not 
create appreciable  neutron skin in $pf$ shell nuclei.  
 This difference of the skin 
thickness gives rise to a  substantial effect on 
 the Q moment.    One of the clear examples of the skin 
effect is seen in $^{50}$Fe where
 the proton deformation is smaller the neutron one, but
 the Q$_{2p}$ moment is larger than Q$_{2n}$ moment. 

\subsection{Hexadecapole deformation in $sd$ and $pf$ shell nuclei}
\begin{figure}[p]
\includegraphics[width=16cm,clip]{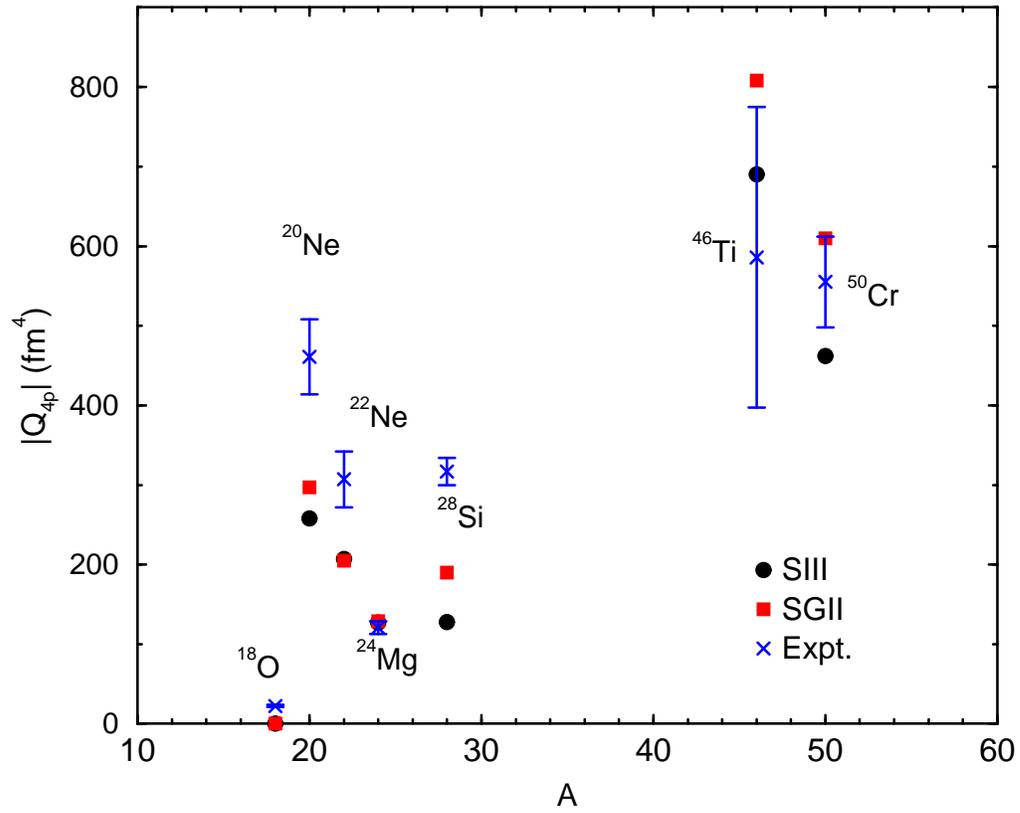}
\caption{\label{fig:q4}
 Proton hexadecapole moment Q$_{4p}$ 
 defined by Eq. (\ref{eq:q4_p}).  The SIII and SGII interactions are used
in the deformed HF calculations with a surface pairing interaction
 (\ref{eq:pair_s}).    Experimental data are taken from refs. 
\cite{Endt93,NNDC}. See the text for details.}  
\end{figure}

Higher multipole deformations 
such as  the hexadecapole deformation were discussed in ref. \cite{JZ84} 
 by the Skyrme HF calculations without pairing correlations. 
It is  an interesting  subject whether the pairing correlations 
are  important or not for the evolution of 
higher multipole deformations than the 
quadrupole.  The proton and neutron hexadecapole deformations of $T$=0 and
$T$=1 nuclei are listed in Tables \ref{tab:Q-T0},  \ref{tab:Q-T1} and 
 \ref{tab:Q-TZm1}.  In $sd$ shell nuclei, a large Q$_4$ in $^{20}$Ne 
and a small Q$_4$ in $^{24}$Mg are expected as the shell filling effect
in prolate deformed potential \cite{BM75a,JZ84}.  The present HF results are 
consistent with the shell filling effect.  The two interactions SIII and
SGII give a large difference in the  Q$_4$ moment of  $^{28}$Si.
In comparison with experimental data in Fig. \ref{fig:q4}, SGII
gives a better prediction in $^{28}$Si. In  $^{32}$S and $^{36}$Ar, 
the calculated  Q$_4$ moments are relatively small compared with 
other  $sd$ shell nuclei, while the shell model gives several times larger
 Q$_4$ moments in these two nuclei \cite{Brown80}. 
 We need more data, for example,
by electron scatterings,  to compare with 
 these calculated results in $^{32}$S and $^{36}$Ar,
 although the proton 
scattering data of $^{32}$S show an indication
 of large  Q$_{4p}$ value \cite{Leo81}.

For $pf$ shell nuclei, the Q$_4$ moments increase drastically from 
$^{44}$Ti and peaked at $^{48}$Cr.  In the heavier $pf$ shell nuclei, 
we can see a large drop in magnitude of  Q$_4$ moment from  $^{48}$Cr 
  to $^{56}$Ni. 
In general, SGII gives larger Q$_4$ moments in $pf$ shell than SIII. 
In $sd$ shell nuclei, the HF results underestimate somewhat the 
empirical values, while the calculations give reasonable values of 
Q$_4$ moment in $pf$ shell nuclei,  $^{46}$Ti and $^{50}$Cr.

\section{SUMMARY}
 We pointed out the strong mass number dependence of quadrupole deformations 
in the nuclei with mass A=(16$\sim$56)
 as a clear  manifestation of the evolution of nuclear deformation 
in atomic nuclei  
by using the deformed HF+BCS  calculations.
The effect in the medium-heavy nuclei is unique to compare  with that in 
rare-earth nuclei since the prolate and  oblate deformations
appear clearly in the ground states  depending 
on the Fermi energy in the deformed single-particle energy levels.
  It is shown  that 
the deformed HF+BCS  model is 
   successful to describe observed Q$_2$ and Q$_4$
  moments of the nuclei with the
isospin T=0 and  T=1 in $sd$ and $pf$ shell in the same level as 
modern shell model calculations.
The isospin symmetry of quadrupole  deformations 
of mirror nuclei with T=1, T$_z=\pm$1 
 is  studied in comparison with the empirical data 
and the shell model calculations.
 It was shown that the value (Q$_{2p}$/Q$_{2n})$/(Z/N) is close to unity 
   in the 
well-deformed nuclei in the middle of the shell, while the values  deviate 
  largely
 from the unity in nuclei near the closed shell both in the 
HF calculations and also in the empirical values. 
The HF results show oblate shapes in some $sd$ shell nuclei
 and the available observed Q$_2(2^+)$ moments of the first excited states 
are consistent with the  theoretical predictions.
 The effect of the pairing interaction on the deformation is 
carefully examined by changing the pairing strength. We point out also
the effect of  large proton skin thickness 
on the Q$_{2}$ moments in the N$\sim$Z nuclei.  

\begin{acknowledgments}
We thank  T. Motobayashi and K. Yamada for showing their data prior 
to a publication. We thank also  M. Honma for providing the results of 
 his shell model
 calculations.  We are benefited from discussions with I. Hamamoto and
 K. Matsuyanagi. This work is supported in part by the Japanese
Ministry of Education, Culture ,Sports, Science  and Technology
  by Grant-in-Aid  for Scientific Research under
 the program number (C(2)) 16540259.
\end{acknowledgments}

\newpage
\noindent {\bf\large Figures }

\newpage

\newpage

\newpage

\newpage

\newpage

\newpage

\newpage

\newpage

\begin{table}[p]
 \squeezetable \caption{ \label{tab:Q-T0}
  Quadrupole deformation parameters $\beta_2$,  quadrupole  moments Q$_2$ 
and hexadecapole moments Q$_4$ in
  $sd$ and $pf$ shell nuclei with the isospin  T=0. The 
deformed HF+BCS calculations are performed with SIII and SGII interactions  
together with the density dependent pairing interaction (\ref{eq:pair_s}).  
  The axial 
symmetry is assumed in the HF calculations.   
  The proton, neutron 
and mass deformation parameters ($\beta_{2p}$,  $\beta_{2p}$ and  $\beta_{2}$)
are obtained at the energy minima of protons, neutron and mass potentials,
 respectively. The proton and neutron quadrupole and hexadecapole moments are 
calculated by using deformed HF wave functions at each energy minima.
  The local minima of the energy surfaces  at low excitation energies are also 
listed.}

\begin{tabular}{l|r|r|r|r|r|r|r|r|r}
\hline
  SIII &$K^{\pi}$& Energy&$\beta_{2p}$&$\beta_{2n}$&$\beta_2$& Q$_{2p}$ & Q$_{2n}$& Q$_{4p}$ & Q$_{4n}$ \\
        &         &  (MeV)&            &            &         &  $(fm^2)$   &  $(fm^2)$     &   $(fm^4)$   &  $(fm^4)$   \\   
\hline
$^{16}O$&$0^+$&0.0& 0.0& 0.0& 0.0& -0.01& -0.01&-0.15&-0.15\\

\hline
$^{20}Ne$&$0^+$       &0.0 & 0.38&  0.34   &  0.38 &39.93&38.87&257.58&243.63 \\
         &$0^+$       &1.98& -0.13& -0.13  &  -0.13&-12.94&-12.55&21.32&20.03 \\
\hline
$^{24}Mg$&$0^+$&0.0 &0.40&0.40&0.40&55.71&54.43 &126.63&123.57\\

\hline
$^{28}Si$& $0^+$ & 0.0& -0.24& -0.24 & -0.24&-39.97&-38.88&127.61&120.73\\
         & $0^+$ &0.86& 0.04& 0.04  & 0.04 & 6.10& 5.92&7.28&6.74\\

\hline
$^{32}S$ &$0^+$ &0.0& 0.20& 0.20& 0.20& 41.62& 40.10&-49.84&-44.64\\
         &$0^+$ &1.54&-0.08&-0.08&-0.08&-16.40&-15.59&-8.82&-7.84\\

\hline
$^{36}Ar$&$0^+$ &0.0& -0.18& -0.18& -0.18& -44.46& -43.01&-47.34&-41.74\\
         &$0^+$ &2.30& 0.09& 0.09& 0.09&22.59&22.01&-15.62&-13.92\\

\hline
$^{40}Ca$&$0^+$ &0.0& 0.01& 0.01& 0.01& 2.72& 2.64&1.17&0.97\\

\hline
$^{44}Ti$&$0^+$ &0.0  &0.15&0.15&0.15&51.73&49.96&508.84&467.90\\
         &$0^+$ &0.20&-0.11&-0.11&-0.11&-35.78&-34.80&155.90&144.70\\
\hline
$^{48}Cr$&$0^+$ &0.0 &0.26& 0.26&0.26&101.80&98.72&785.70&737.63\\
         &$0^+$ &3.55&-0.11& -0.11&-0.11&-41.17&-39.86&40.45&35.71\\

\hline
$^{52}Fe$&$0^+$ &0.0 &0.20& 0.20&0.20&90.72&88.19&-27.00&-16.71\\
         &$0^+$ &3.04&-0.08&-0.08&-0.08&-34.16&-33.22&-32.77&-29.07\\

\hline
$^{56}Ni$&$0^+$ &0.0 &0.01& 0.01&0.01&4.78&4.63&-12.13&-11.04\\

\hline
\end{tabular}
\end{table}
\begin{table}
 \squeezetable 
\begin{tabular}{l|r|r|r|r|r|r|r|r|r}
\hline
 SGII   &$K^{\pi}$& Energy&$\beta_{2p}$&$\beta_{2n}$&$\beta_2$& Q$_{2p}$ & Q$_{2n}$& Q$_{4p}$ & Q$_{4n}$ \\
        &         &  (MeV)&            &            &         &  $(fm^2)$   &  $(fm^2)$     &   $(fm^4)$   &  $(fm^4)$   \\ 
\hline
$^{16}O$&$0^+$&0.0& 0.0& 0.0& 0.0& -0.01& -0.01&-0.17&-0.17\\

\hline
$^{20}Ne$&$0^+$       &0.0 & 0.41&  0.41& 0.41 &43.38&42.12&297.09&279.87 \\
         &$0^+$       &2.69& -0.15& -0.15& -0.15&-14.81&-14.33&29.61&27.60\\
\hline
$^{24}Mg$&$0^+$&0.0 &0.42&0.42&0.42&56.88&55.42&129.01&125.44\\

\hline
$^{28}Si$& $0^+$ & 0.0& -0.29& -0.29 & -0.29&-49.12&-47.58&190.08&178.53\\
         & $0^+$ &1.97& 0.10& 0.10  & 0.10 & 15.11& 14.67&21.40&20.01\\

\hline
$^{32}S$ &$0^+$ &0.0& 0.14& 0.13& 0.13& 27.04& 25.83&-27.16&-24.14\\

\hline
$^{36}Ar$&$0^+$ &0.0& -0.17& -0.17& -0.17& -42.21& -40.66&-64.86&-57.59\\
         &$0^+$ &1.63& 0.09& 0.09& 0.09&22.45&21.85&-15.14&-13.26\\

\hline
$^{40}Ca$&$0^+$ &0.0& 0.0& 0.0& 0.0& 0.03& 0.04&0.05&0.03\\

\hline
$^{44}Ti$&$0^+$ &0.0  &0.19&0.19&0.19&65.48&62.80&823.01&750.34\\
         &$0^+$ &0.68&-0.10&-0.10&-0.10&-32.54&-31.59&140.74&130.16\\
\hline
$^{48}Cr$&$0^+$ &0.0 &0.31& 0.31&0.31&123.69&119.18&1057.68&985.81\\
         &$0^+$ &4.46&-0.12& -0.12&-0.12&-44.72&-43.38&52.80&50.20\\

\hline
$^{52}Fe$&$0^+$ &0.0 &0.23& 0.23&0.23&99.0&95.85&49.32&55.63\\
         &$0^+$ &3.72&-0.12&-0.12&-0.12&-51.23&-49.57&106.35&98.11\\

\hline
$^{56}Ni$&$0^+$ &0.0  &0.0  & 0.0 &0.0  &-0.01 &-0.0&-60.41&-54.23\\

\hline

\end{tabular}
\end{table}

\newpage

\newpage

\begin{table}[p]
 \squeezetable \caption{\label{tab:Q-T1}
  Quadrupole deformation parameters $\beta_2$,  quadrupole  moments Q$_2$ 
and hexadecapole moments Q$_4$ in  $sd$ and $pf$ shell nuclei with the isospin  T=1 and T$_z=-1$. The 
deformed HF+BCS calculations are performed with SIII and SGII interactions 
together with the density-dependent pairing interaction. 
See the caption to Table~\ref{tab:Q-T0} for details.}

\begin{tabular}{l|r|r|r|r|r|r|r|r|r}
\hline
   SIII &$K^{\pi}$& Energy&$\beta_{2p}$&$\beta_{2n}$&$\beta_2$& Q$_{2p}$ & Q$_{2n}$& Q$_{4p}$ & Q$_{4n}$ \\
        &         &  (MeV)&            &            &         &  $(fm^2)$   &  $(fm^2)$     &   $(fm^4)$   &  $(fm^4)$   \\

\hline
$^{18}Ne$&$0^+$ &0.0 & 0.01& 0.01& 0.01&1.16&0.43&-2.87&-0.49 \\

\hline
$^{22}Mg$&$0^+$&0.0 &0.40&0.401&0.40&54.65&42.00&175.85&207.73\\
         &$0^+$&3.16&-0.15&-0.16&-0.16&-20.26&-15.58&31.97&29.20\\

\hline
$^{26}Si$&$0^+$&0.0 &0.21&0.28&0.24&34.88&37.53&59.15&49.84\\
         &$0^+$&0.33&-0.21&-0.21&-0.21&-34.86&-27.63&90.01&64.14\\

\hline
$^{30}S$ &$0^+$ &0.0& 0.10& 0.09& 0.09& 19.17& 15.01&1.98&10.49\\

\hline
$^{34}Ar$&$0^+$&0.0&-0.16 &-0.16&-0.16&-39.94&-32.24&-36.68&-12.87\\
         &$0^+$&0.21&0.13&0.16&0.14&31.38&33.45&-28.93&-26.07\\

\hline
$^{38}Ca$&$0^+$ &0.0& 0.01& 0.01& 0.01& 2.20& 2.58&0.81&1.35\\

\hline
$^{42}Ti$&$0^+$ &0.0& 0.01& 0.01& 0.01& 3.83& 2.08&-10.42&-3.26\\

\hline
$^{46}Cr$&$0^+$ &0.0& 0.21& 0.19& 0.20& 83.97& 65.13&744.56&639.78\\
         &$0^+$ &1.33& -0.11& -0.11& -0.11&-40.82&-35.16&90.35&112.74\\
\hline
$^{50}Fe$&$0^+$ &0.0  &0.23&0.24&0.23&100.19&91.12&390.70&499.18\\
         &$0^+$ &2.87&-0.10&-0.11&-0.11&-45.39&-41.56&50.28&42.44\\

\hline
$^{54}Ni$&$0^+$ &0.0 &0.11& 0.14&0.12&52.96&60.37&86.04&30.38\\

\hline
\end{tabular}

\begin{tabular}{l|r|r|r|r|r|r|r|r|r}
\hline
   SGII &$K^{\pi}$& Energy&$\beta_{2p}$&$\beta_{2n}$&$\beta_2$& Q$_{2p}$ & Q$_{2n}$& Q$_{4p}$ & Q$_{4n}$ \\
        &         &  (MeV)&            &            &         &  $(fm^2)$   &  $(fm^2)$     &   $(fm^4)$   &  $(fm^4)$   \\   

\hline
$^{18}Ne$&$0^+$ &0.0 & 0.01& 0.06& 0.01&1.40&0.35&-0.42&-0.20\\

\hline
$^{22}Mg$&$0^+$&0.0 &0.41&0.41&0.40&54.74&42.01&176.14&213.04\\
         &$0^+$&3.63&-0.20&-0.19&-0.19&-25.62&-18.73&57.89&45.24\\

\hline
$^{26}Si$&$0^+$&0.0 &0.24&0.318&0.27&38.47&41.49&73.97&66.93\\
         &$0^+$&0.18&-0.24&-0.24&-0.24&-39.77&-30.80&118.14&85.27\\

\hline
$^{30}S$ &$0^+$ &0.0& -0.09& -0.11&-0.10&-17.90&-16.85&-3.26&9.88\\
         &$0^+$ &0.0& 0.03& 0.04&0.04&6.73&6.76&4.93&5.33\\

\hline
$^{34}Ar$&$0^+$&0.0  &-0.13 &-0.12&-0.13&-31.81&-23.86&-42.35&-26.89\\
         &$0^+$&0.26&0.11  &0.13 &0.12 &23.73 &21.36&-16.79&-18.03\\

\hline
$^{38}Ca$&$0^+$ &0.0& 0.01& 0.01& 0.01& 2.10& 2.68&0.81&1.30\\

\hline
$^{42}Ti$&$0^+$ &0.0& 0.0& -0.00& -0.00& 0.01& 0.0&-4.22&-1.32\\

\hline
$^{46}Cr$&$0^+$ &0.0& 0.248& 0.221& 0.235& 96.03& 73.77&914.29&787.16\\
         &$0^+$ &2.13& -0.13& -0.13& -0.13&-48.46&-41.11&149.33&171.21\\

\hline
$^{50}Fe$&$0^+$ &0.0 &0.25&0.26&0.25&108.03&98.0&394.96&523.95\\
         &$0^+$ &3.70&-0.13&-0.13&-0.13&-54.28&-48.21&106.15&74.09\\

\hline
$^{54}Ni$&$0^+$ &0.0 &0.13& 0.16&0.15&61.17&69.13&60.88&8.37\\
         &$0^+$ &0.53 &-0.07& -0.08&-0.07&-31.51&-31.40&4.79&-9.09\\

\hline
\end{tabular}
\end{table}

\newpage

\begin{table}[p]
 \squeezetable \caption{\label{tab:Q-TZm1}
Quadrupole deformation parameters $\beta_2$,  quadrupole  moments Q$_2$ 
and hexadecapole moments Q$_4$  in
  $sd$ and $pf$ shell nuclei with the isospin  T=1 and T$_z=1$. The 
deformed HF+BCS calculations are performed with SIII and SGII interactions  
together with the density-dependent pairing interaction. 
See the caption to Table~\ref{tab:Q-T0} for details.}

\begin{tabular}{l|r|r|r|r|r|r|r|r|r}
\hline
   SIII &$K^{\pi}$& Energy&$\beta_{2p}$&$\beta_{2n}$&$\beta_2$& Q$_{2p}$ & Q$_{2n}$& Q$_{4p}$ & Q$_{4n}$ \\
        &         &  (MeV)&            &            &         &  $(fm^2)$   &  $(fm^2)$     &   $(fm^4)$   &  $(fm^4)$   \\   
\hline
$^{18}O$&$0^+$&0.0& -0.01& -0.01& -0.01& -0.51& -1.00&-0.13&-0.10\\

\hline
$^{22}Ne$&$0^+$       &0.0 & 0.39& 0.38  & 0.39 &41.38&51.15&206.75&159.93  \\
         &$0^+$       &3.08& -0.16& -0.15  &  -0.16&-15.99&-19.55&30.64&30.17 \\
\hline
$^{26}Mg$&$0^+$&0.0 &0.29&0.22&0.25&39.64&35.41&59.63&64.73\\
         &$0^+$&0.31& -0.21& -0.21& -0.21& -28.33& -33.78&66.75&84.20\\

\hline
$^{30}Si$& $0^+$ & 0.0& 0.08& 0.09 & 0.08&13.99&16.55&9.92&3.22\\

\hline
$^{34}S$ &$0^+$ &0.0& -0.16& -0.16& -0.16& -33.34& -38.41&-13.58&-29.72\\
         &$0^+$ &0.29&0.16&0.12&0.14&32.59&28.67&-27.90&-24.61\\

\hline
$^{38}Ar$&$0^+$ &0.0& -0.01& -0.01& -0.01& -3.49& -1.78&-0.39&-0.44\\

\hline
$^{42}Ca$&$0^+$ &0.0& -0.01& -0.01& -0.01& -2.19& -3.85&-0.88&-2.47\\

\hline
$^{46}Ti$&$0^+$ &0.0  &0.20&0.22&0.21&70.40&84.65&689.82&688.03\\
         &$0^+$ &1.29&-0.11&-0.11&-0.11&-36.09&-39.59&121.36&86.17\\
\hline
$^{50}Cr$&$0^+$ &0.0 &0.24&0.23&0.23&93.70&97.15&461.74&302.57\\
         &$0^+$ &2.81 &-0.10&-0.10&-0.11&-42.66&-43.97&43.84&46.20\\

\hline
$^{54}Fe$&$0^+$ &0.0 &0.12& 0.09&0.11&52.72&43.51&17.18&59.43\\

\hline
$^{58}Ni$&$0^+$ &0.0 &0.11& 0.14&0.13&56.09&74.64&238.35&343.72\\
         &$0^+$ &0.17&-0.09&-0.10&-0.09&-42.38&-50.42&115.25&94.07\\

\hline
\end{tabular}

\begin{tabular}{l|r|r|r|r|r|r|r|r|r}
\hline
   SGII &$K^{\pi}$& Energy&$\beta_{2p}$&$\beta_{2n}$&$\beta_2$& Q$_{2p}$ & Q$_{2n}$& Q$_{4p}$ & Q$_{4n}$ \\
        &         &  (MeV)&            &            &         &  $(fm^2)$   &  $(fm^2)$     &   $(fm^4)$   &  $(fm^4)$   \\   
\hline
$^{18}O$&$0^+$&0.0& 0.01& 0.01& 0.01& 0.37& 1.27&-0.19&-0.27\\

\hline
$^{22}Ne$&$0^+$       &0.0 & 0.41& 0.40  & 0.41 &43.19&53.21&225.14&171.92\\
         &$0^+$       &3.61& -0.18& -0.19&  -0.18&-18.39&-23.47&43.48&47.64 \\
\hline
$^{26}Mg$&$0^+$&0.0 &0.32&0.24&0.28&42.59&37.48&71.51&73.05\\
         &$0^+$&0.16& -0.24& -0.24& -0.24& -31.71& -38.40&89.68&109.92\\

\hline
$^{30}Si$& $0^+$ & 0.0& -0.02&-0.02 & -0.02&-3.44&-3.57&-5.25&-6.15\\

\hline
$^{34}S$ &$0^+$ &0.0& -0.13& -0.14& -0.14& -26.93& -32.46&-27.97&-36.41\\
         &$0^+$ &0.25&0.14&0.11&0.12&27.51&25.70&-26.42&-19.14\\

\hline
$^{38}Ar$&$0^+$ &0.0& 0.01& 0.01& 0.01& 2.69& 2.11&0.38&0.15\\

\hline
$^{42}Ca$&$0^+$ &0.0& 0.01& 0.01& 0.01& 2.0& 3.83&0.64&1.78\\

\hline
$^{46}Ti$&$0^+$ &0.0  &0.22&0.25&0.24&76.50&92.14&807.53&796.05\\
         &$0^+$ &2.07&-0.11&-0.11&-0.11&-35.86&-39.73&139.57&105.34\\

\hline
$^{50}Cr$&$0^+$ &0.0 &0.27&0.26&0.26&105.20&108.17&609.74&419.16\\
         &$0^+$ &3.65&-0.14&-0.14&-0.14&-53.17&-56.62&94.77&123.07\\

\hline
$^{54}Fe$&$0^+$ &0.0 &0.15& 0.11&0.13&62.84&50.50&23.90&82.27\\

\hline
$^{58}Ni$&$0^+$ &0.0 &-0.11& -0.12&-0.11&-53.14&-59.25&173.44&118.77\\
         &$0^+$ &0.02&0.12&0.14&0.13&56.64&73.71&195.14&238.98\\

\hline
\end{tabular}
\end{table}

\begin{table}[p]
 \squeezetable \caption{\label{tab:q2-moment}
Quadrupole moments Q$_2$(2$^+$) 
 of the first excited 2$^+$ states in $sd$ and $pf$ shell nuclei.
The unit is $e\cdot fm^2$. Shell model values of  $sd$ shell nuclei are 
 taken from ref. \cite{Wil71}, while those of $pf$ shell nuclei 
are taken from  ref. \cite{Honma04}. 
 Experimental data are taken from the compilation of ref. \cite{Rag89}.}

\begin{tabular}{l|r|r|r|r }
\hline
 nucleus& SIII & SGII& shell model & Expt.\\
\hline

$^{18}$O&  0.15  & -0.10  & -2.0 & -3.9 $\pm$ 0.9   \\
$^{20}$Ne & -11.4   & -12.4  & -12.1 & -23 $\pm$ 3 \\
$^{22}$Ne &  -12.3  & -12.3  & -13.6&  -19 $\pm$ 4  \\
$^{24}$Mg& -15.9   &  -16.3 & -15.0  & -18  $\pm$2     \\
$^{28}$Si & 11.4   & 14.0  & 14.3  &  16$\pm$ 3   \\
$^{30}$Si & -4.0   & 0.97  & -6.6   & -5 $\pm$ 6  \\
$^{32}$S &   -11.9 & -7.7  &  -13.6  & -15.4 $\pm$ 2.0   \\
$^{34}$S & 9.5   & 7.7  & 6.7   &4  $\pm$ 3  \\
$^{36}$Ar& 13.4   & 12.1  & 14.3   & 11 $\pm$  6 \\
$^{42}$Ca & 0.63   &  0.57 & $ 1.90$   &  -19$\pm$ 8   \\
$^{46}$Ti &  -20.1  & -21.9  &$-11.1$   &  -21$\pm$6   \\
$^{50}$ Cr& -26.7   & -30.0  & -26.4  & -36 $\pm$7   \\
$^{54}$Fe & -15.1    & -17.9  & -22.6   & -5 $\pm$14  \\
$^{58}$Ni &  -16.0  & -16.2   &  -2.4  & -10 $\pm$ 6 \\
\hline
\end{tabular}
\end{table}

\end{document}